\documentclass[prc,aps,amsfonts,amsmath,superscriptaddress,floatfix,twocolumn]{revtex4}
\usepackage{graphicx,float} 
\usepackage{epsfig} 
\usepackage{rotating}
\usepackage{mwe}   
\usepackage{subfig}
\begin{document}

\title{Derivation of the M$_n$/M$_p$ ratio in exotic nuclei}

\author{E. Khan}
\affiliation{IJCLab, Universit\'e Paris-Saclay, CNRS/IN2P3, 91405 Orsay Cedex, France}

\begin{abstract} 
A generalized formula is provided, to calculate the M$_n$/M$_p$ ratio of the multipole transition matrix elements, in the framework of the so-called phenomenological analysis. 
It takes into account the possible difference between the neutron and proton radii and diffuseness, which can occur, especially in exotic nuclei.
The validity domain of the original Bernstein formula is discussed, in the case of the proton scattering probe at few tens of MeV. The largest discrepancies are obtained for very neutron-rich nuclei (N/Z$\gtrsim$1.6) or when the electromagnetic deformation parameter is larger than the proton scattering one. The reduction of the statistical error bars, and the study of very neutron-rich nuclei at facilities of exotic beams of  new generation, should favor the use of the generalized Bernstein formula. 
\end{abstract}
 


\date{\today}

\maketitle

\section{Introduction}

Information, extracted from nuclear excitations can be related to two main quantities: the energy spectrum, and the excitation probability. The latter is usually decomposed into a neutron and a proton contribution, leading to the neutron and proton matrix elements, M$_n$ and M$_p$, respectively \cite{ber81,ber83,bm69}. The proton one can be directly determined from electromagnetic measurements, such as Coulomb excitation. However, determining the neutron matrix element is more delicate. In practice, another probe is used, such as proton scattering, which involves both protons and neutrons of the nucleus of interest. It is then necessary to combine the electromagnetic and proton scattering data, to extract the M$_n$ and M$_p$ values. With the advent of exotic beam facilities, the M$_n$/M$_p$ ratio can be now analyzed along several isotopic chains (see e.g. \cite{cal16,kun19,cor18}), and provide information about e.g. modification of the magicity. 

In practice, a microscopic analysis is considered as the most relevant method to extract the M$_n$ and M$_p$ values. The angular distributions are predicted using optical, and transition potentials, involving neutron and proton transition densities. When both the B(EL) and the angular distribution data are well described, the transition densities are validated, providing the M$_n$ and M$_p$ values. Of course, this is an ideal case, since the microscopic transition densities may have to be renormalized in order to describe the data \cite{kha01,bec06}. The microscopic approach may also yield some uncertainties, due to the choice of the parameters of both the optical and transition potentials (nucleon-nucleus parameterization, double folding, etc.). The uncertainty on the transition densities should also be considered, such as the validity of assuming them spherically symmetric. All in all, it may not be efficient to use a fully detailed microscopic analysis, in order to ultimately calculate an integrated quantity, such as M$_n$ and M$_p$.

Another way to analyse the data, is the so-called phenomenological analysis. In this approach, the M$_n$/M$_p$ ratio is directly determined, assuming a macroscopic behavior for the neutrons and protons of the nucleus of interest. The respective distributions of neutron and proton fluids are deformed, due to the excitation \cite{bm69}. In this method, a phenomenological optical potential is used, such as the 
Becchetti-Greenlees \cite{bec69} or the Koning-Delaroche \cite{kd03} parameterizations, in the case of proton scattering. The corresponding deformation parameter $\beta_{n,p}$ is determined by normalizing the predicted angular distribution to the data (see e.g. \cite{cot01,ken92,kha00,kha01}). In order to further extract the M$_n$/M$_p$ ratio, the so-called Bernstein formula \cite{ber81,ber83,cot01} is used, which allows to relate this ratio to the two deformation lengths, extracted from the electromagnetic and a second probe (named hereafter F), such as proton scattering:  

\begin{equation}
\frac{M_n}{M_p}=\frac{b_p}{b_n}\left[\frac{\delta^F}{\delta_p}\left(1+\frac{N}{Z}\frac{b_n}{b_p}\right)-1\right]
\label{eq:bern}
\end{equation}

$\delta$ is the deformation length corresponding to the electromagnetic ($\delta_p$) or second ($\delta^F$) probe, and relates the transition to the optical potentials, as recalled in the next section.
The Bernstein formula also involves b$_{n,p}$, the interaction strength between the external field and a neutron or a proton of the nucleus, respectively. 

However, the validity of formula (\ref{eq:bern}) may be limited: it assumes identical neutron and proton radii \cite{ber81,ber83}, which is not the case, in e.g. exotic nuclei. It has also been known 
to provide questionable results in several isotopic chains, such as Ar, S and O isotopes \cite{kha00,kha01}, although the large statistical error bars, due to the low rate of events with exotic beams, preclude for making strong statements. 

With the advent of new exotic beams facilities, such error bars are expected to be reduced, and more and more neutron-rich nuclei shall be studied. In this framework, and in order to provide a phenomenological extraction of the M$_n$ and M$_p$ values with an extended validity, it could be relevant to generalize the Bernstein formula to the case of exotic nuclei. A by-product is also to provide a derivation of the original Bernstein formula, since the author could not find it in the literature. A generalized Bernstein formula could help in 
providing a quick method to extract the M$_n$/M$_p$ ratio, including exotic nuclei and isotopic chains.

In section II, the generalized Bernstein formula is derived. Additional hypotheses necessary to reach the original one are discussed, and differences are also analyzed. Section III provides a phenomenological input for the neutron and proton radii and diffuseness, which are necessary to compute, in a convenient way, the generalized Bernstein formula. 
The comparison of the M$_n$/M$_p$ ratios, between the original and the generalized Bernstein formulae, is then discussed in the case of 50 MeV/A proton scattering, on the S, Ar and Oxygen isotopic chains.

\section{The generalized Bernstein formula}

\subsection{Derivation}

In this section, a derivation of the generalized Bernstein formula is provided. The additional hypotheses to reach the original formula are then discussed. 

 An optical potential U, corresponding to a given probe F interacting with a nucleus, can be decomposed as
 \begin{equation}
 U=U_n+U_p
 \label{eq:u}
 \end{equation}
 where U$_{n,p}$ is the contribution to the optical potential from the neutron(protons) of the nucleus \cite{sat83}.

The transition potential $\Delta$U is generated by the deformation of the nucleus, during the excitation, due to the probe F \cite{sat83}. The radius R$_0$ is changed by a quantity, the deformation length $\delta^F\equiv$R$_0\beta$, where $\beta$ is the deformation parameter, such that:

\begin{equation}
R=R_0+\delta^F
\end{equation}

 and hence
  
 \begin{equation}
 U(R)=U(R_0+\delta^F)\equiv U(R_0)+ \Delta U 
\end{equation}

with

\begin{equation}
 \Delta U \simeq \delta^F \left.\frac{dU}{dr}\right|_{r=R_0}
 \label{eq:duu}
\end{equation}

In order to provide a generalization of the Bernstein formula, let us derive the relation between the transition potential and its optical potential, assuming a typical Woods-Saxon shape with potential depth U$_0$ and diffuseness a, for the latter:

 \begin{equation}
U(r)=\frac{U_0}{1+e^\frac{r-R_0}{a}}
\end{equation}

This leads to:
\begin{equation}
\left.\frac{dU}{dr}\right|_{r=R_0}=-\frac{U_0}{4a}
\end{equation}

and therefore, with Eq. (\ref{eq:duu}):

 \begin{equation}
 \Delta U\simeq -\frac{R_0\beta U_0}{4a}
  \label{eq:du}
\end{equation}

Eq. (\ref{eq:u}) implies  $\Delta$U=$\Delta$U$_n$+$\Delta$U$_p$, which gives,  with Eq. (\ref{eq:du}):
 \begin{equation}
 \frac{U_0\delta^F}{a}= \frac{ U_p\delta_p}{a_p}+\frac{U_n\delta_n }{a_n}
 \label{eq:tp}
 \end{equation}
 
 where U$_{n,p}$ now corresponds to the depth of the neutron and proton contribution to the optical potential, and $\delta_{n,p}\equiv R_{n,p}\beta_{n,p}$. It should be noted that Eq. (\ref{eq:tp}) is a generalization of equation (14.124) of \cite{sat83}, taking into account the radius and diffuseness of the potentials.
 
 Eqs (\ref{eq:u}) and (\ref{eq:tp}) allow the calculation of the ratio of the magnitudes of the neutron and proton contributions to the optical potential:
 
 \begin{equation}
 \frac{U_n}{U_p}=\frac{\frac{\delta^F}{\delta_p}\frac{a_p}{a}-1}{\frac{\delta_n}{\delta_p}\frac{a_p}{a_n}-\frac{\delta^F}{\delta_p}\frac{a_p}{a}}
 \label{eq:int}
\end{equation}

This ratio can also be written, using the interaction strength b$_{n,p}$ of a single neutron (proton) of the nucleus with the external probe, as

\begin{equation}
 \frac{U_n}{U_p}= \frac{Nb_n}{Zb_p}
 \label{eq:ub}
\end{equation}

Finally, in the framework of the macroscopic model \cite{ber83,ken92}, the multipole matrix transition elements M$_{n,p}$ are related to the deformation parameters by:
\begin{equation}
\frac{M_n}{M_p}=\frac{N}{Z}\frac{\beta_n}{\beta_p}
\label{eq:mb}
\end{equation}

Eq. (\ref{eq:mb}) shows that when M$_n$/M$_p$ is larger than N/Z, the amplitude of the neutron oscillation, during the excitation, is larger than the proton one, and vice-versa.

Eq. (\ref{eq:mb}) allows to write the $\delta_n/\delta_p$ ratio of Eq. (\ref{eq:int}) as a function of  M$_n$/M$_p$, and, together with Eq. (\ref{eq:ub}), provides a generalized expression of the Bernstein formula:
\begin{equation}
\frac{M_n}{M_p}=\frac{b_p}{b_n}\frac{a_n}{a_p}\frac{R_p}{R_n}\left[\frac{\delta^F}{\delta_p}\frac{a_p}{a}\left(1+\frac{N}{Z}\frac{b_n}{b_p}\right)-1\right]
\label{eq:bgen}
\end{equation}

where a$_{n,p}$ and R$_{n,p}$ are recalled to be the neutron and proton diffuseness and radii, respectively.

\subsection{Reduction to the original Bernstein formula}

To recover the original Bernstein formula, it is necessary to assume identical shapes for the total, neutron and proton optical potentials, namely a=a$_n$=a$_p$ and R$_n$=R$_p$. In this case, Eq. (\ref{eq:bgen}) becomes the usual Bernstein formula (\ref{eq:bern}). 

It should be noted that Eq. (\ref{eq:bgen}) still assumes identical deformation parameters for the potentials and the densities: $\beta$ of Eq (\ref{eq:du}) is assumed to be 
the same as the one of  Eq (\ref{eq:mb}). It also assumes the validity of the macroscopic model (Eq (\ref{eq:mb})).  

Compared to the original one (Eq. (\ref{eq:bern})), the generalized Bernstein formula (\ref{eq:bgen}) is more relevant to study the (M$_n$/M$_p$)/(N/Z) ratio, over isotopic chains involving exotic nuclei, which can have qualitatively different neutron and proton densities. 

The specific case $\delta^F/\delta_p$=1 is useful to illustrate the difference between the original and the generalized Bernstein formulae: the former gives M$_n$/M$_p$=N/Z, i.e. $\beta_n$=$\beta_p$, from Eq. (\ref{eq:mb}), whereas the latter still allows for a different deformation parameter between neutrons and protons, due to the possible differences in the shape of the neutron and proton densities.

\section{Results}

\subsection{Phenomenological implementation}

The generalized Bernstein formula (\ref{eq:bgen}) involves the proton and neutron radii R$_{n,p}$ and diffuseness a$_{n,p}$. A phenomenological parameterization for these quantities is provided
in the present section, in order to use Eq. (\ref{eq:bgen}) in a convenient way. 

In the case of radii, the following parameterization \cite{war98,wan01} is used, based on the comparison with extensive relativistic mean field (RMF) microscopic calculations. 
It should be noted that this parameterization has been also validated by comparison with the experimental data, and includes the effect of the isospin dependence.  

For light nuclei (A$<40$) \cite{wan01}:

\begin{equation}
R_{n,p}=r_{0}A^{1/3}+r_{1}+r_{2}\left(\frac{N-Z}{A}\right)+r_{3}\left(\frac{N-Z}{A}\right)^2
\label{eq:r1}
\end{equation}

where the r$_i$ parameters are given in Table \ref{tab:rnp}, both for the neutron and the proton cases. We have taken here, the parameterization which is valid for most of the nuclei, but not for the halo ones. The binding energy parameter of the formulae (9) and (10) of Ref. \cite{wan01} is taken to 8 MeV. We have checked that, taking different values of binding energy, such as 6 MeV, changes the final M$_n$/M$_p$ value by less that 5\%. Finally, it should be noted that the values of the r$_i$ parameters of Ref. \cite{wan01} are mutliplied by a $\sqrt{5/3}$ factor, due to the relation between the radii and the rms radii (see e.g. Eq. (10) of Ref. \cite{war98}).

\begin{center}
\begin{table}[ht]
\begin{tabular}{c|c|c|c|c||c|c|c|c|}
 \cline{2-8}
                 & \multicolumn{4}{c||}{A $<$ 40} & \multicolumn{3}{c|}{A $\geq$40} \\
   \cline{2-8}
     & r$_0 $ & r$_1$&r$_2$ &r$_3$ &r$_4$ &r$_5$ &r$_6$ \\
\hline
   R$_n$ &1.02 & 0.75 & 0.46&1.08  &1.18 &3.30 &0.29 \\
\hline
   R$_p$ & 1.03 & 0.79& -1.07&0.82 &1.24 &-0.80 &-0.19 \\
\hline
\end{tabular}
\caption{Parameters (in fm) of the neutron and proton radii phenomenological formulae (\ref{eq:r1}) and (\ref{eq:r2}).}
\label{tab:rnp}
\end{table}
\end{center}

For heavier (A$\geq$40) nuclei, the parameterization, which best corresponds to the RMF results, is \cite{war98}:
\begin{equation}
R_{n,p}=r_{4}A^{1/3}+\frac{r_5+r_6\left(N-Z\right)}{A^{2/3}}   
\label{eq:r2}
\end{equation}

where the r$_i$ parameters, for the neutron and proton cases, are also given in Table \ref{tab:rnp}.

Diffuseness can be related to the separation energy, using \cite{gam85}:

\begin{equation}
a_{n,p}\simeq\frac{\hbar}{2\sqrt{2m_{n,p}S_{n,p}}}
\label{eq:as}
\end{equation}

 It is then relevant to use a parameterization for the separation energy, which has been determined in order to best describe the experimental values \cite{vog01}:
 
\begin{equation}
S_{n,p}=(s_1+s_2A^{1/3})\left(\frac{N}{Z}\right)^\alpha-s_3+\frac{s_4}{A^{1/2}}-\eta s_5-s_6\frac{Z}{A^{1/3}}
\label{eq:s}
\end{equation}

where the s$_i$ and $\alpha$ parameters are given in Table \ref{tab:sa}. The $\eta$ parameter, in Eq. (\ref{eq:s}), corresponds to the shell correction effect: $\eta$=0 if the corresponding N or Z value is lower than 28,  $\eta$=1 if this value is between 29 and 50, $\eta$=2 for a value between 51 and 82, etc. 

\begin{center}
\begin{table}[ht]
\begin{tabular}{c|c|c|c|c|c|c|c|c|}
\cline{2-8}
   &s$_1$ & s$_2$ & s$_3$&s$_4$  &s$_5$ &s$_6$&$\alpha$  \\
\hline
   S$_n$ &6.29 & 3.43 & 5.85 &10.59  &1.51 & 0 & -1 \\
\hline
   S$_p$ & 13.83 & 0.64& 3.56 &12.14 &1.32 &0.98 & 1 \\
\hline
\end{tabular}
\caption{Parameters (in MeV for s$_1$ to s$_6$) for the phenomenological formula (\ref{eq:s}) of the neutron and proton separation energies. The value of s$_4$ in S$_{n,p}$ should be changed by a minus sign in the case of an odd value of N or Z, respectively (pairing effect).}
\label{tab:sa}
\end{table}
\end{center}

To summarize, Eqs. (\ref{eq:r1}) and (\ref{eq:r2}) allow the calculation of the R$_p$/R$_n$ factor of the generalized Bernstein formula (\ref{eq:bgen}). Eq. (\ref{eq:as}) provides the values of a$_n$ and a$_p$, using Eq. (\ref{eq:s}). Finally, the diffuseness a of the optical potential in Eq. (\ref{eq:bgen}) is taken to

\begin{equation}
 a=0.7 fm,
 \label{eq:aop}
 \end{equation}
which is a value close to the one used for the diffuseness of the real part of the phenomenological optical potentials, such as the Becchetti and Greenlees \cite{bec69} or the Koning and Delaroche \cite{kd03} ones.

Finally, for a convenient use of Equation (\ref{eq:bgen}), Table \ref{tab:bnbp} recalls the usual phenomenological values of the b$_n$/b$_p$ ratio \cite{ber81,ber83,cot01}. 

\begin{center}
\begin{table}[ht]
\begin{tabular}{c|c|c}
 Probe  & Energy &  b$_n$/b$_p$ \\
 \hline
 Electromagnetic & All & 0   \\
 Protons  & 10-50 MeV &  3 \\
 Protons  & 1 GeV &  0.95 \\
 Neutrons  & 10-50 MeV &  1/3 \\
   $\alpha$ & All &  1 \\
    $\pi^+$  & 160-200 MeV &  1/3 \\
    $\pi^-$  & 160-200 MeV &  3 \\
\end{tabular}
\caption{Ratio of the neutron to proton interaction probability, of a probed nucleus, as a function of the probe \cite{ber81}}
\label{tab:bnbp}
\end{table}
\end{center}

\subsection{Results and comparison with the original Bernstein formula}

Eq (\ref{eq:bgen}) can now be used, with the paremeterizations and values provided in section III.A for  
R$_{n,p}$, a$_{n,p}$, b$_{n,p}$ and a. In order to evaluate the impact, on the M$_n$/M$_p$ ratio, of the generalization of the Bernstein formula, we consider the case of proton scattering at about 50 MeV/nucleon: the O, S and Ar isotopic chains have been studied both microscopically and phenomenologicaly \cite{kha00,kha01} in this case. 

\begin{figure}[tb]
\scalebox{0.35}{\includegraphics{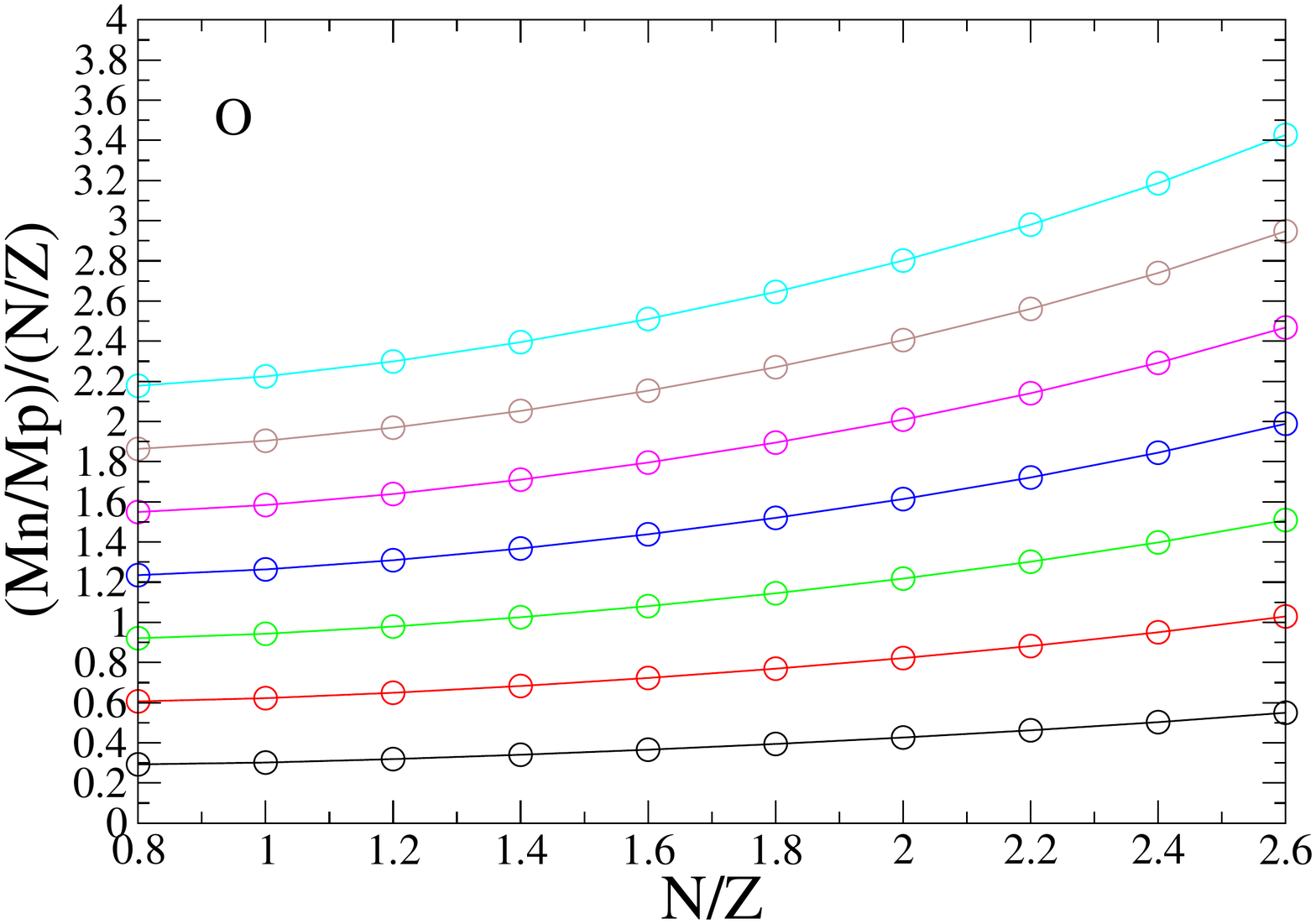}}
 \caption{(M$_n$/M$_p$)/(N/Z) ratio calculated with Eq (\ref{eq:bgen}) and the parameterisations of radii and diffuseness from sect III.A (circles), for $\delta^F$/$\delta_p$ ratios ranging
 from 0.5 to 2 (from bottom to top), with step of 0.25, in the case of Oxygen isotopes.}
 \label{fig:mnmpO}
\end{figure}

\begin{figure}[tb]
\scalebox{0.35}{\includegraphics{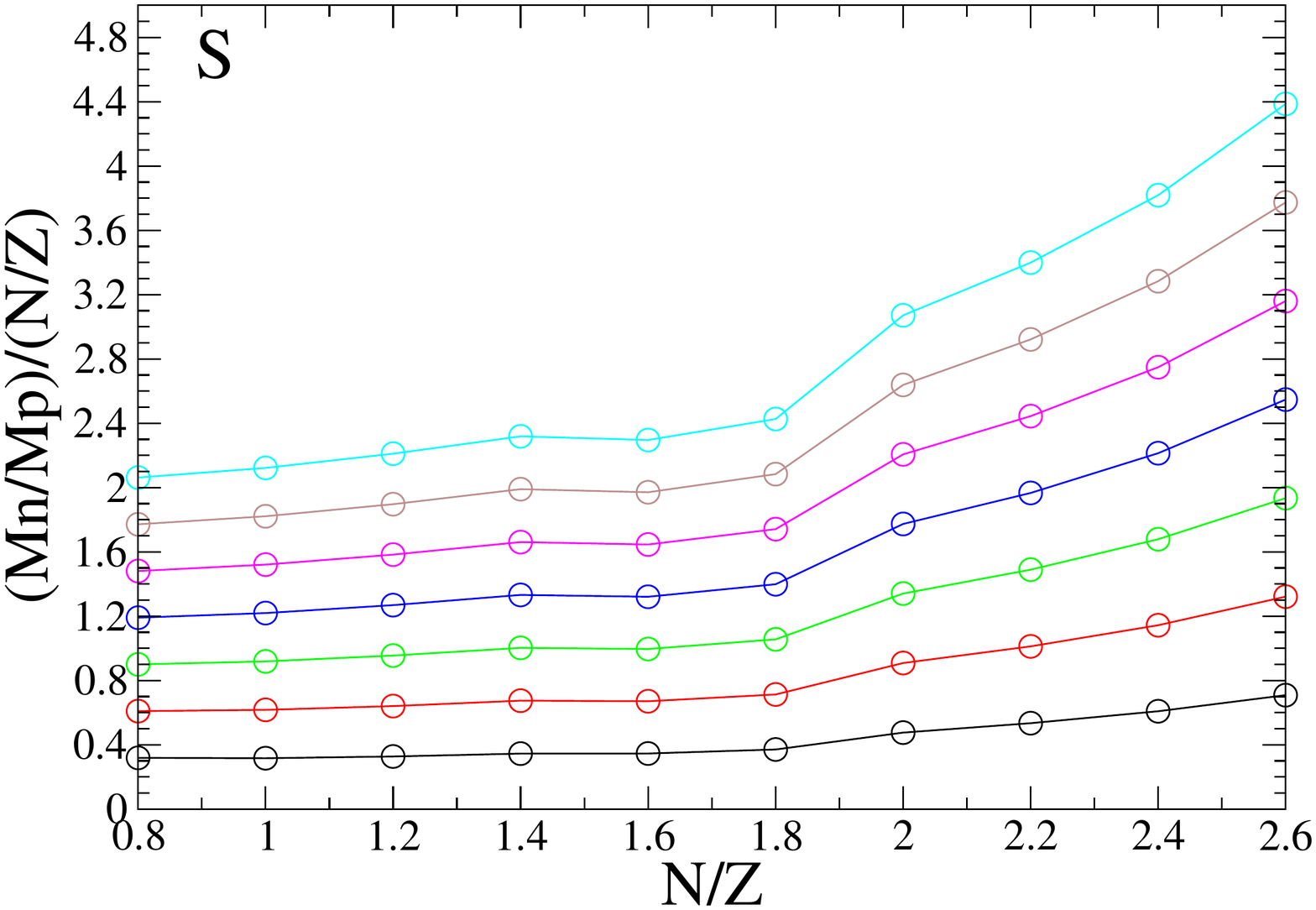}}
 \caption{Same as Fig. \ref{fig:mnmpO} for Sulfur isotopes}
 \label{fig:mnmpS}
\end{figure}

\begin{figure}[tb]
\scalebox{0.35}{\includegraphics{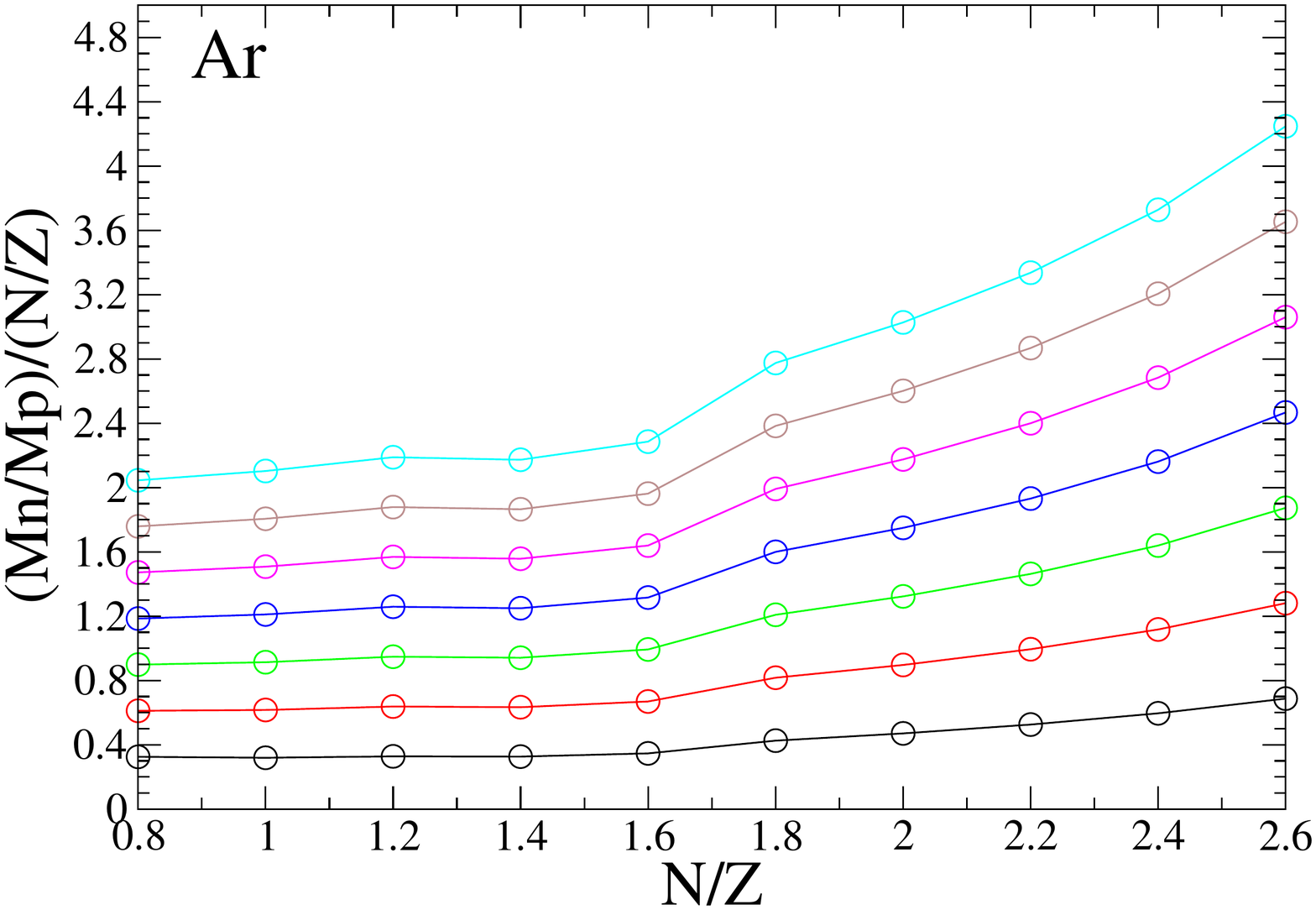}}
 \caption{Same as Fig. \ref{fig:mnmpO} for Argon isotopes}
 \label{fig:mnmpAr}
\end{figure}

\begin{figure}[tb]
\scalebox{0.35}{\includegraphics{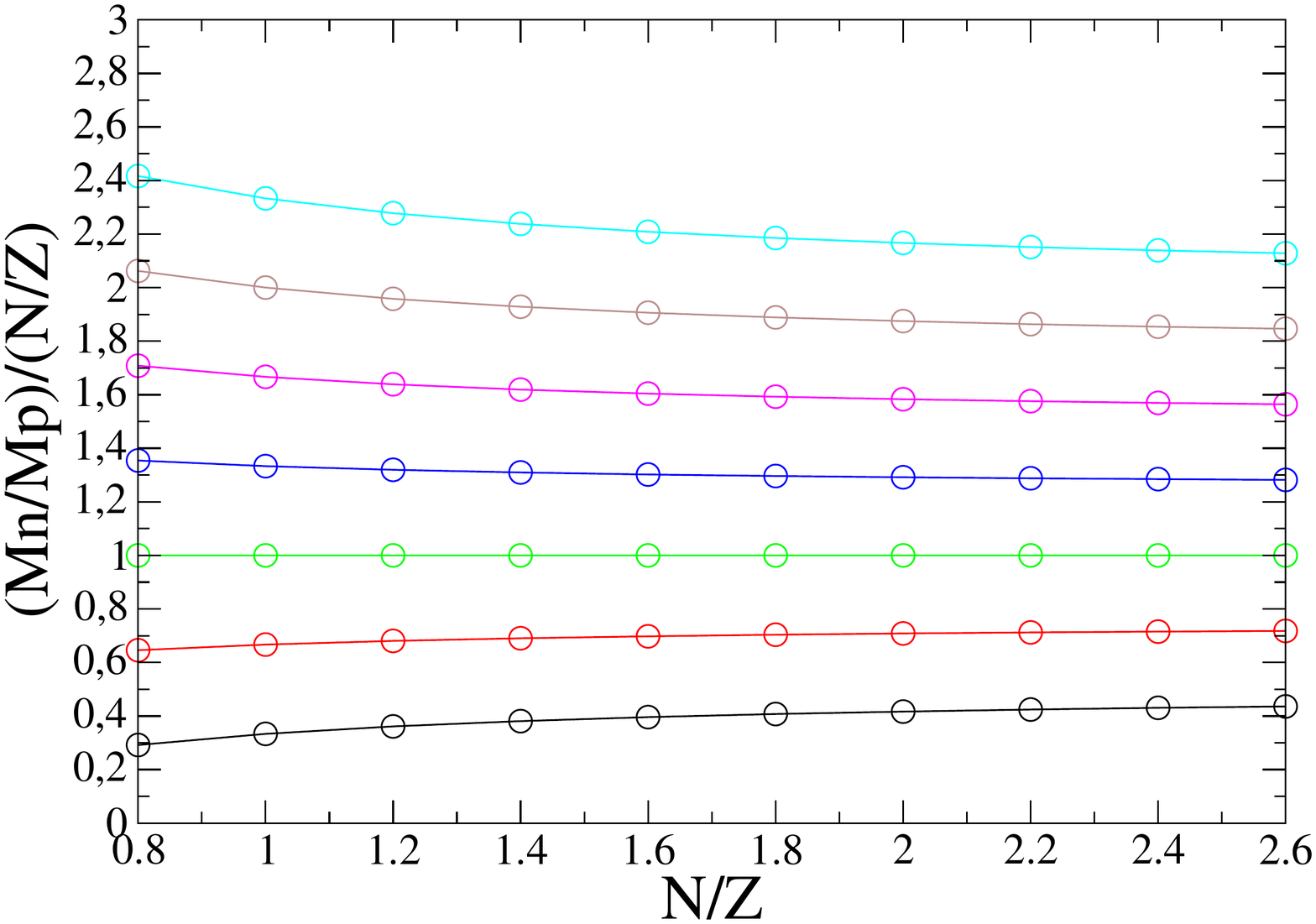}}
 \caption{(M$_n$/M$_p$)/(N/Z) ratio calculated with the original Bernstein formula (\ref{eq:bern}) (circles), for $\delta^F$/$\delta_p$ ratios ranging
 from 0.5 to 2 (from bottom to top), with step of 0.25. }
 \label{fig:mnmp}
\end{figure}

In a first step, the behaviors of the original and the generalized Bernstein formulae are compared, as a function of the $\frac{\delta^F}{\delta_p}$ ratio of the deformations lengths, which are usually determined experimentally. Figures \ref{fig:mnmpO}, \ref{fig:mnmpS} and \ref{fig:mnmpAr} display the (M$_n$/M$_p$)/(N/Z) ratios, obtained from Eq (\ref{eq:bgen}), for various $\frac{\delta^F}{\delta_p}$ values, in the case of proton scattering at about 50 MeV/nucleon, for the O, S and Ar isotopic chains, respectively. For comparison, the same ratio using the original Bernstein formula is also plotted in Fig. \ref{fig:mnmp}. In this latter case,
the calculated M$_n$/M$_p$ ratio does not depend on the radii and diffuseness of the considered nucleus (see Eq. (\ref{eq:bern})). 

A striking pattern is that the trend of the  M$_n$/M$_p$ ratios, which are rather constant for N/Z $\lesssim$ 1.6 and increases for more neutron rich nuclei with the generalized formula, is not well described by the original Bernstein 
formula: in this last case, a smooth behavior is obtained, without any strong increase of the M$_n$/M$_p$ ratio for neutron-rich nuclei. Moreover,  Fig. \ref{fig:mnmpO}, \ref{fig:mnmpS} and \ref{fig:mnmpAr} exhibit a different behavior between the 3 isotopic chains. This emphases the limitation of the original Bernstein formula, for which an identical behavior is predicted for all nuclei, for given N/Z and $\frac{\delta^F}{\delta_p}$ values.

\begin{figure}[tb]
\scalebox{0.35}{\includegraphics{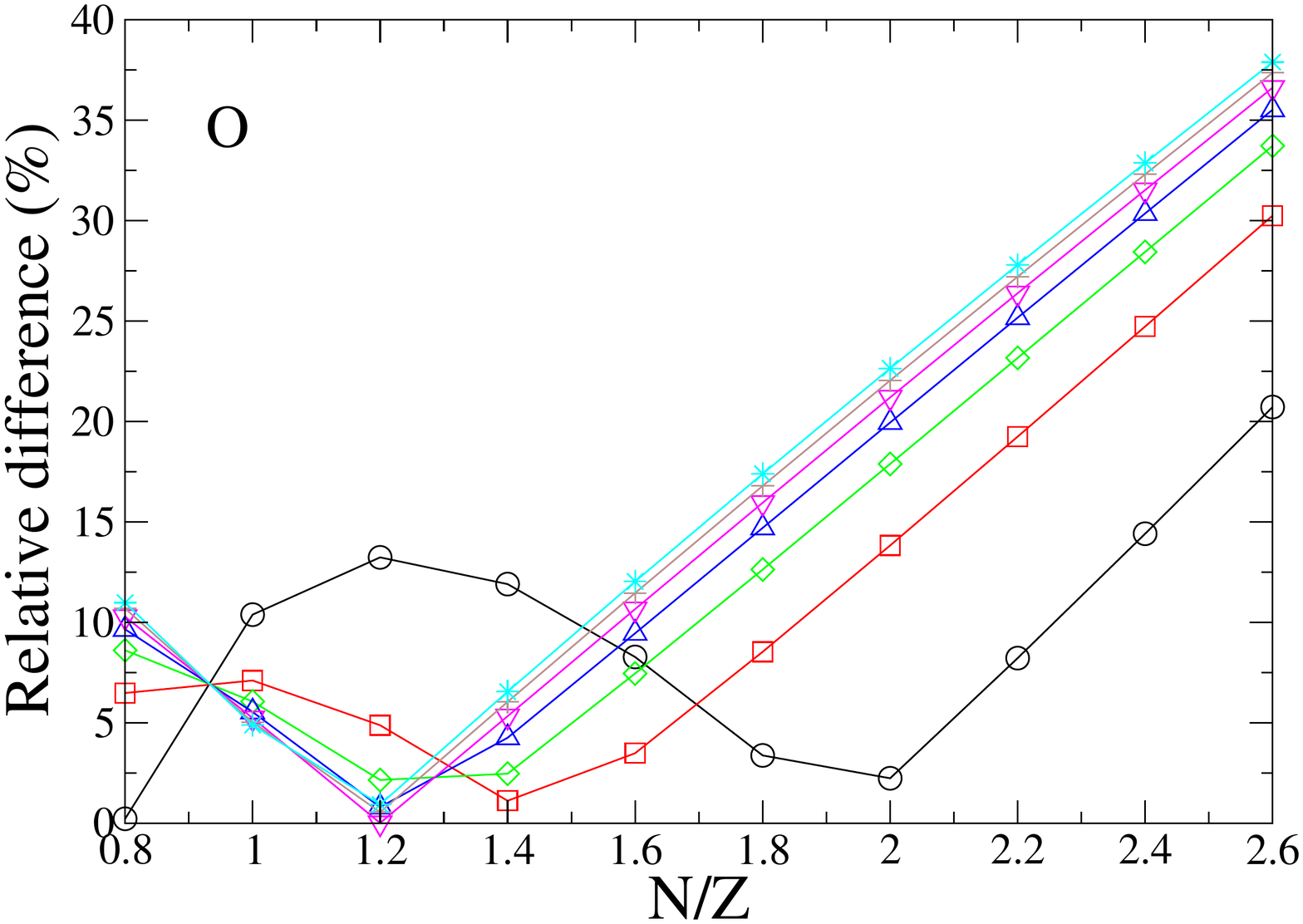}}
 \caption{Relative difference of the M$_n$/M$_p$ ratios calculated with Eqs (\ref{eq:bern}) and (\ref{eq:bgen}) (symbols), for $\delta^F$/$\delta_p$ ratios ranging
 from 0.5 to 2 (from bottom to top), with step of 0.25, in the case of Oxygen isotopes. See Fig. \ref{fig:mnmpO} for the color code.}
 \label{fig:relO}
\end{figure}

\begin{figure}[tb]
\scalebox{0.35}{\includegraphics{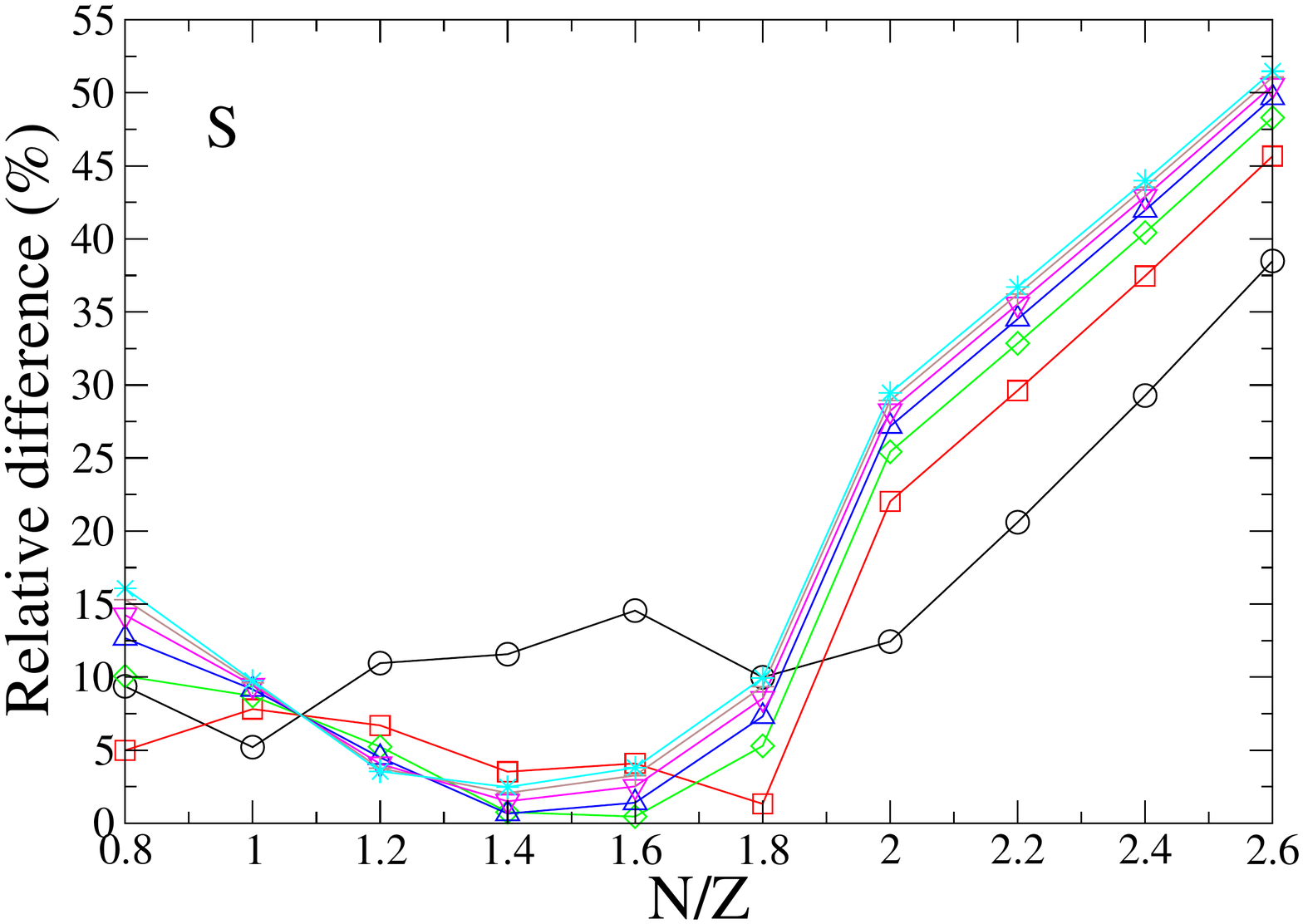}}
 \caption{Same as Fig. \ref{fig:relO} for Sulfur isotopes.}
 \label{fig:relS}
\end{figure}

\begin{figure}[tb]
\scalebox{0.35}{\includegraphics{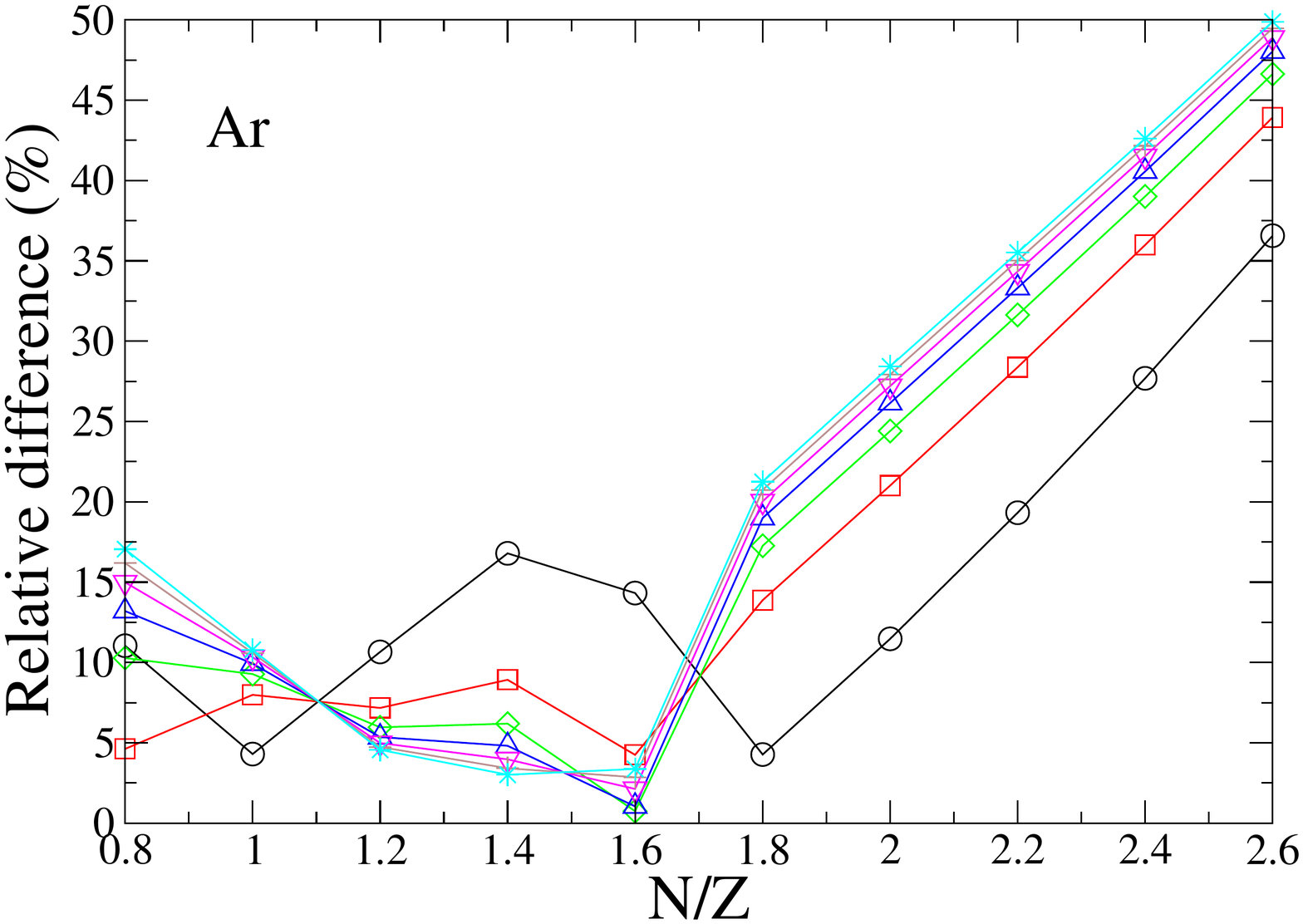}}
 \caption{Same as Fig. \ref{fig:relO} for Argon isotopes.}
 \label{fig:relAr}
\end{figure}

In order to study more quantitatively the difference between the original and generalized Bernstein formulae, Fig. \ref{fig:relO}, \ref{fig:relS} and \ref{fig:relAr} display their relative difference on the three isotopic chains of Oxygen, Sulfur and Argon, respectively. The largest differences are obtained for very neutron rich nuclei, with N/Z ratio larger than 1.6 or 1.8, such as $^{48}$S and beyond. This result is expected, because the original Bernstein formula assumes identical shapes for proton and neutron densities (radius and diffuseness), which is not the case for such neutron-rich nuclei. A more unexpected limitation of the original Bernstein formula impacts nuclei with N/Z=0.8, such as $^{14}$O, $^{28}$S or
 $^{32}$Ar. In these slightly proton-rich nuclei, the generalized Bernstein formula leads to variations between 10\% and 20\% on the M$_n$/M$_p$ ratio, compared to the original one. 
 Finally, excited states for which $\delta_p$ is larger than deformation length $\delta^F$ of the probe, are also significantly not well described by the original Bernstein formula, namely for $\delta^F$/$\delta_p$=0.5 or 0.75.


\subsection{Results from the Oxygen, Sulfur and Argon data}

In a second step, we now compare the M$_n$/M$_p$ values extracted from the data measured on the first 2$^+$ states in the Oxygen, Sulfur and Argon isotopic chains, namely with the electromagnetic probe, and the proton scattering one at around 50 MeV per nucleon \cite{kha00,kha01,bec06}. The M$_n$/M$_p$ ratio is extracted in three different ways: i) using the original Bernstein formula (\ref{eq:bern}), from a macroscopic analysis using the Becchetti and Greenlees parameterization of the optical potential \cite{bec69}. The deformation parameter is obtained by normalizing the theoretical inelastic  angular distribution to the data (see e.g. Eq. (\ref{eq:du})). ii) the M$_n$/M$_p$ ratio is also extracted by using the generalized Bernstein formula (\ref{eq:bgen}), with the same method. It should be noted that, in these phenomenological approaches, the same radii have been taken for the deformation length of both probes \cite{ken92}, to extract their deformation parameters. iii) the M$_n$/M$_p$  values obtained from the microscopic calculation of \cite{kha00,kha01,bec06} is also recalled. In this approach, microscopic transition densities from BCS+QRPA calculations are used with the Jeukenne-Lejeune-Mahaux (JLM) proton-nucleus optical potential \cite{jlm}. Assuming the shapes of these microscopic densities are relevant, their magnitudes are determined as follows: the proton one, $\delta\rho_p$, from the B(E2), and the neutron one $\delta\rho_n$, using the proton scattering data. The obtained normalization values are then used to microscopically calculate  the M$_n$/M$_p$ ratio, with:

\begin{equation}\label{eq:mn}
M_{n,p} = \int \delta\rho_{n,p}(r) r^{L+2}dr~,
\end{equation}

where L is the multipolarity of the excitation (here L=2).

In principle, the better the phenomenological analysis, the closer to the microscopic values, which are usually considered as more sound. However, the microscopic analysis also has its own limitations, such as using spherical densities for deformed nuclei, or the limitations of the JLM approach, with a necessary renormalisation of the imaginary part of the optical potential \cite{pet93}. 

The (M$_n$/M$_p$)/(N/Z) values for the Sulfur isotopes are displayed in Table {\ref{tab:mnmpS}. The errors bars are determined from the experimental error bars of both 
the electromagnetic and proton scattering data. The two phenomenological analysis and the microscopic one provides the same trend: an increasing (M$_n$/M$_p$)/(N/Z) ratio with increasing neutron number, except in the case of $^{36}$S. In this nucleus, the
microscopic analysis finds a decrease of the neutron contribution to the excitation, which can be understood from the N=20 shell closure. The generalized Bernstein formula gives a somewhat smaller value than the one from the original formula, but it is not possible to make a strong assessment, because of the error bars. On average, the generalized Bernstein formula gives values closer to the microscopic one, than the original formula. In the case of $^{38}$S, both macroscopic values are significantly larger than the microscopic one, although compatible within their error bars. It should be recalled that the microscopic analysis is not 
a straight data value, it also comes with an analysis process and its own limitations.

\onecolumngrid
\begin{center}
\begin{table}[h]
\begin{tabular}{|c||c|c|c|c|c|c|}
$\frac{(M_n/M_p)}{(N/Z)}$ & \textbf{$^{30}S$}   & \textbf{$^{32}S$}  & \textbf{$^{34}S$} &
  \textbf{$^{36}S$}  & \textbf{$^{38}S$}  & \textbf{$^{40}S$} 
\\ \hline
  Original & 0.93 $\pm$ 0.20& 0.95 $\pm$ 0.11 & 0.91 $\pm$ 0.11 & 1.13 $\pm$
  0.27 & 1.50 $\pm$ 0.30 & 1.25 $\pm$ 0.25
\\ \hline
 Generalized  & 0.85 $\pm$ 0.18&  0.87 $\pm$ 0.10 & 0.88 $\pm$ 0.13 & 1.08  $\pm$ 0.34 & 1.53 $\pm$ 0.33 & 1.26 $\pm$ 0.31
\\ \hline
  Micro & 0.88 $\pm$ 0.21& 0.94 $\pm$ 0.16 & 0.85 $\pm$ 0.23 & 0.65 $\pm$
  0.18 & 1.09 $\pm$ 0.29 & 1.01 $\pm$ 0.27 
\\ 
\end{tabular}
\caption{\label{tab:mnmpS}}
(M$_n$/M$_p$)/(N/Z) ratios for the Sulfur isotopes calculated with the original Bernstein formula (\ref{eq:bern}), the generalized one (\ref{eq:bgen}) and the microscopic analysis of Ref. \cite{kha01}.
\end{table}
\end{center}
\twocolumngrid

In the case of the Argon isotopes (Table {\ref{tab:mnmpar}), the same trend is observed, namely the generalized Bernstein formula gives M$_n$/M$_p$ values usually closer to the microscopic one, than with the original Bernstein formula, but the error bars do not allow to make a strong difference. This shows that, in principle, the generalized Bernstein formula is more accurate 
than the original one, in the sense that the predicted values are closer to the microscopic one. More accurate measurements, with the upgrade of both detection setups and exotic beam
facilities should provide a reduction of the errors bars, and confirm the usefulness of the generalized Bernstein formulae.

\onecolumngrid
\begin{center}
\begin{table}[h]
\begin{tabular}{|c||c|c|c|c|c|}
 $\frac{(M_n/M_p)}{(N/Z)}$ & \textbf{$^{34}Ar$} & \textbf{$^{36}Ar$}
 & \textbf{$^{40}Ar$} & \textbf{$^{42}Ar$} &
 \textbf{$^{44}Ar$} 
 \\ \hline
  Original & 1.18 $\pm$ 0.25& 1.57 $\pm$ 0.25  & 0.75 $\pm$ 0.05 & 1.14 $\pm$
  0.17 & 1.26 $\pm$ 0.16 
\\ \hline
 Generalized  & 1.06  $\pm$ 0.21&  1.40 $\pm$ 0.25 & 0.68 $\pm$ 0.07 & 1.05 $\pm$ 0.23 & 1.21 $\pm$ 0.19
\\ \hline
  Micro & 0.99 $\pm$ 0.17& 1.41 $\pm$ 0.50 & 0.68 $\pm$ 0.21 & 1.12 $\pm$
  0.26 & 1.13 $\pm$ 0.27 
\\
\end{tabular}
\caption{\label{tab:mnmpar}}
Same as Table \ref{tab:mnmpS}, for Argon isotopes.
\end{table}
\end{center}
\twocolumngrid

In the case of Oxygen isotopes (Table {\ref{tab:mnmpo}), the three analysis are still compatible within the error bars. They all describe a decrease of the neutron contribution, from $^{20}$O to $^{22}$O, which is interpreted as a N=14 shell closure \cite{bec06}. The original Bernstein values are at similar level of agreement with the microscopic one, than the generalized formula. It could be, that for such light nuclei, the parameterizations described in section II are less accurate. It could also be, that the microscopic analysis is not much more accurate than the phenomenological one, in the case of Oxygen isotopes. In any case, the difference between the original and the generalized Bernstein formulae is expected to be larger in the case of proton-rich isotopes, such as $^{14}$O, as it can be seen from Fig. \ref{fig:relO}, especially if the the electromagnetic deformation parameter is larger than the neutron one.

\begin{center}
\begin{table}[h]
\begin{tabular}{|c||c|c|c|}
 $\frac{(M_n/M_p)}{(N/Z)}$ & $^{18}$O & $^{20}$O
 & $^{22}$O
 \\ \hline
  Original & 1.05 $\pm$ 0.13& 2.35 $\pm$ 0.37  & 1.28 $\pm$ 0.51
\\ \hline
 Generalized  & 1.05 $\pm$ 0.14 &  2.60 $\pm$ 0.42 & 1.48$\pm$ 0.63
\\ \hline
  Micro & 0.88 $\pm$ 0.19& 2.17 $\pm$ 0.53 & 1.4 $\pm$ 0.5
\\
\end{tabular}
\caption{\label{tab:mnmpo}}
(M$_n$/M$_p$)/(N/Z) ratios for the Oxygen isotopes calculated with the original Bernstein formula (\ref{eq:bern}), the generalized one (\ref{eq:bgen}) and the microscopic analysis of Ref. \cite{kha00,bec06}
\end{table}
\end{center}

More generally, the error bars for all these isotopes are between 10\% and 30\%: it is difficult to disentangle the results of the generalized Bernstein formula, from the ones of the original 
formula, as shown on Figs. \ref{fig:relO}-\ref{fig:relAr}. Although the original Bernstein formula does not describe qualitatively the behavior of the M$_n$/M$_p$ ratio, it ironically describes it quantitatively, because of the still large error bars on this ratio.

\section{Conclusion}

A generalized formula is provided, to derive the M$_n$/M$_p$ ratio from the deformation parameters obtained from the electromagnetic and a second probe, such as proton scattering at few tens of MeV. It also allows, in the same way, to rederive the original so-called Bernstein formula: the generalized formula depends on the neutron and proton radii, which were assumed to be identical in the original derivation,
as well as for neutron and proton diffuseness.

In order to easily use the generalized formula, phenomenological parameterizations of radii and diffuseness are recalled, which are valid over the nuclear chart. This implementation allows the comparison with the original Bernstein formula, in the case of proton scattering at few tens of MeV per nucleon. The difference is usually less than 10\%, except for proton-rich (N$\lesssim$0.8Z), very neutron-rich nuclei (N$\gtrsim$1.5Z), or for cases where the electromagnetic deformation parameter is larger than the proton scattering one. The original Bernstein formula does not 
qualitatively describe the behavior of the M$_n$/M$_p$ ratio of the generalized formula.

The present approach has been discussed on the example of the Sulfur, Argon and Oxygen data. On average, a better agreement with the microscopic analysis 
is observed in the case of the generalized Bernstein formula. However, due to current errors bars, and in the case of not too exotic nuclei, the original and Bernstein 
M$_n$/M$_p$ ratios are compatible. However, with the increase of the detection capabilities, errors bars are expected to be reduced, as well as more exotic nuclei shall become accessible. 
This shall allow a better assessment of  the improvement brought by the generalized formula.

The validity of the present generalized Bernstein formula, has been studied in the case of proton scattering at few tens of MeV per nucleon. It would be relevant to also study it for other probes, in a future work,
although there is a priori no reason, that would make this formula less valid in the case of other probes. 

\begin{acknowledgments}
The author thanks Y. Blumenfeld for fruitful remarks and reading the manuscript,
and F. Mercier for fruitful discussions.
\end{acknowledgments}


\end{document}